\documentclass[aps,pra,twocolumn,showpacs]{revtex4}
\usepackage{stmaryrd}
\usepackage{amsfonts}
\usepackage{mathrsfs}
\usepackage{amsmath}
\usepackage{amscd}
\usepackage{graphicx}
\usepackage{booktabs}
\usepackage{leftidx}
\begin{document}
\title{Controllable single-photon transport between remote coupled-cavity arrays}
\author{Wei Qin$^{1,2}$}%\protect\thanks{Corresponding author:  qinwei09@tsinghua.org.cn}
\author{Franco Nori$^{1,3}$}
\address{$^1$ CEMS, RIKEN, Wako-shi, Saitama 351-0198, Japan\\
$^2$ School of Physics, Beijing Institute of Technology, Beijing 100081, China\\
$^3$ Department of Physics, University of Michigan, Ann Arbor, Michigan 48109-1040, USA}

\begin{abstract}
We develop a new approach for controllable single-photon transport between two remote one-dimensional coupled-cavity arrays, used as quantum registers, mediated by an additional one-dimensional coupled-cavity array, acting as a quantum channel. A single two-level atom located inside one cavity of the intermediate channel is used to control the long-range coherent quantum coupling between two remote registers, thereby functioning as a quantum switch. With a time-independent perturbative treatment, we find that the leakage of quantum information can in principle be made arbitrarily small. Furthermore, our method can be extended to realize a quantum router in multi-register quantum networks, where single-photons can be either stored in one of the registers or transported to another on demand. These results are confirmed by numerical simulations.
\end{abstract}
\maketitle
%%%%%%%%%%%%%%%%%%%%%%%%%%%%%%%%%%%%%%%%%%%%%%%%%%%%%%%%%%%%%%%%%%%%%%%%%%%%%%%%%%%%%%%%%
\section{introduction}
\label{se:section1}
Quantum networks are fundamental for quantum information science \cite{net1,net2}. An elementary quantum network is composed of spatially-separated quantum nodes for quantum information manipulation and storage, with these nodes connected by quantum channels for quantum information distribution \cite{net3}. Thus, the implementation of such a quantum network relies upon the ability to realize the reliable transport of quantum states through these quantum channels. To this end, in the form of flying qubits, photons serve as an optimal choice for carrying information for long-distance quantum communications \cite{photon1,photon2,photon3,photon4,photon5}. Another approach to connect distant qubits is to utilize solid-state systems \cite{solid}. Such solid-state devices include electron spins of nitrogen-vacancy (NV) colour centers in diamond \cite{NV1,NV2,NV3,NV4,NV5,NV6}, nuclear spins in nuclear magnetic resonance (NMR) \cite{nuclear1,nuclear2}, flux qubits in superconductors \cite{flux1,flux2,flux3,flux4}, cold atoms in optical lattices \cite{cold1,cold2,cold3}, and even magnons in ferromagnets \cite{mag1,mag2,mag3,mag4}. Moreover, in recent decades, coupled-cavity arrays (CCAs) are currently being explored, for example, in superconducting transmission line resonators \cite{STLR1,STLR2,STLR3,STLR4,STLR5}, photonic crystal resonators \cite{PCR1,PCR2,PCR3} or toroidal microresonators \cite{TM1,TM2,TM3}. The CCAs offer an inherent advantage because each cavity can be individually addressed. Indeed, both coherent optical information storage and transport can be achieved in such arrays, and at the same time the need for an external interface between the quantum register and the quantum channel is eliminated because they use the same fundamental hardware.

In addition to simulating quantum many-body phenomena \cite{MB1,MB2,MB3}, these CCAs also demonstrate promising applications in controlling photon coherent transport by using single controllable two-level or three-level atoms \cite{STLR3,CCA_switch1,CCA_switch2,CCA_switch3,CCA_switch4,CCA_switch5,CCA_switch6,CCA_switch7}. Photons are transmitted or reflected based upon tuning the photon-atom scattering. In this case, the atom behaves as a quantum switch. Despite having been extensively studied \cite{ext1,ext2,ext3,ext4,ext5,ext6,ext7}, prior work on the coherent transport of photons has typically focused on the nearby CCAs via the photon-atom scattering. However, in order to carry out quantum network operations, information needs to be controllably transported between distant quantum registers. Thus, a detailed understanding of controllable quantum channels which could connect these distant registers is of both fundamental and practical importance.

 Here we theoretically introduce a novel method for controllable coherent transport of single-photons upon making use of a CCA, one cavity of which contains a two-level atom, as a quantum channel to connect two remote CCAs as quantum registers. The key element underlying our method is that the atom is harnessed to control the long-range coherent interaction between the two boundary registers. Specifically, within the weak-coupling regime, the registers are resonantly coupled by a specific collective eigenmode of the bare channel, yielding an effective photon transport channel, such that time evolution results in a swap operation of the two registers. However, when this eigenmode is coupled to the atom, it will be dressed and split into two dressed modes. If the splitting between the two dressed modes is significantly detuned from the registers, photons will thus be reflected back and, as a result, the time evolution functions as an identity operation. Furthermore, we directly extend this approach to the case of multi-register quantum networks, where a single-photon can be redirected to different registers at will, in an analogous manner to a quantum router. As opposed to previous work, the proposed model can be applicable to controlling the coherent transport of a single-photon being in an arbitrary quantum state between two remote quantum registers over an arbitrarily long range.

%%%%%%%%%%%%%%%%%%%%%%%%%%%%%%%%%%%%%%%%%%%%%%%%%%%%%%%%%%%%%%%%%%%%%%%%%%%%%%%%%%%%%%%%%%%%%%%%%%%%%%%%%%%%%%%%%%%%%%%%%%%%%%%%%%%%%%%%%%%%%%%%%%%%%%%%%%%%%%%%%%%%%%%%%%%%%%%%%%%%%%%%%
%%%%%%%%%%%%%%%%%%%%%%%%%%%%%%%%%%%%%%%%%%%%%%%%%%%%%%%%%%%%%%%%%%%%%%%%%%%%%%%%%%%%%%%%%%%%%%%%%%%%%%%%%%%%%%%%%%%%%%%%%%%%%%%%%%%%%%%%%%%%%%%%%%%%%%%%%%%%%%%%%%%%%%%%%%%%%%%%%%%%%%%%%
\section{physical model and controllable transport of single-photons}

%%%%%%%%%%%%%%%%%%%%%%%%%%%%%%%%%%%%%%%%%%%%%%%%%%%%%%%%%%%%%%%%%%%%%%%%%%%%%%%%%%%%%%%%%%%%
\begin{figure}[!ht]%[tpb]
\begin{center}
\includegraphics[width=8.5cm,angle=0]{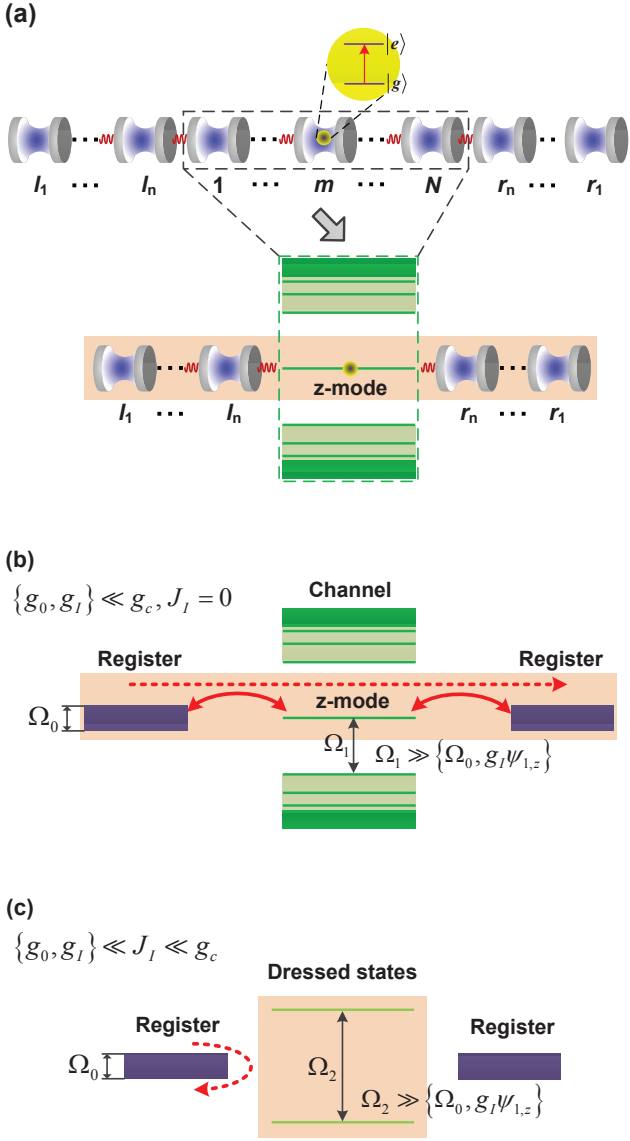}
\caption{(color online) (a) A 1D CCA of having $N$ cavities and a two-level atom is employed as a quantum channel to connect two distant quantum registers composed of two identical 1D CCAs, each contains $n$ cavities. In the limit of $\left\{g_{0},g_{I},J_{I}\right\}\ll g_{c}$, the full dynamics can be reduced to an effective model, only involving the two boundary registers, the atom and the zero-energy mode of the bare channel. (b) Effective coupling configuration in the no-atom case of $J_{I}=0$. By ensuring $\{g_{0},g_{I}\}\ll g_{c}$, the boundary registers are resonant with a single boson mode ($k=z$) while the large detunings are eliminated, so that unitary evolution will result in a swap operation between the two registers. (c) Effective coupling configuration in the single-atom case of $\{g_{0},g_{I}\}\ll J_{I}\ll g_{c}$. Owing to the large detunings between the registers and the dressed states, such registers are decoupled from the intermediate channel. The incoming photon is thus reflected off this channel, and the quantum state of the photon will remain unchanged after time evolution.}\label{fig1}
\end{center}
\end{figure}
%%%%%%%%%%%%%%%%%%%%%%%%%%%%%%%%%%%%%%%%%%%%%%%%%%%%%%%%%%%%%%%%%%%%%%%%%%%%%%%%%%%%%%%%%%%%%%

The basic idea is to use two identical one-dimensional (1D) CCAs to enact quantum registers connected by a quantum channel consisting of an additional CCA and a two-level atom, shown schematically in Fig. \ref{fig1}(a). Let $\hat{c}_{i}^{\dag}$ ($i=1,\cdots,N$) be the creation operator of the $i$th cavity of the channel, and $\hat{c}_{l_{j}/r_{j}}^{\dag}$ ($j=1,\cdots,n$) be that of the $j$th cavity of the left or right register, assuming that all cavities have a common frequency $\omega$. The atom, characterized by a ground state $|g\rangle$ and an excited state $|e\rangle$, is embedded in the $m$th cavity of the channel and is resonantly coupled to the mode of this cavity with strength $J_{I}$. We assume that the intrachannel coupling $g_{c}$ is fixed, and the intraregister coupling $$g_{j}=g_{0}\sqrt{j\left(2n+1-j\right)}/2,$$ with $g_{0}$ being a constant, is non-uniform \cite{nonuniform1,nonuniform2}, which reveals that each register supports a linear spectrum of $$\lambda_{q}=g_{0}\left(2q-n-1\right),$$ where $q=1,\cdots,n$. In a frame rotating at $\omega$, the Hamiltonian governing the total system is
\begin{eqnarray}\label{T_Hami}
\hat{H}_{T}&=&\sum_{d=l,r}\sum_{j=1}^{n-1}g_{j}\left(\hat{c}_{d_{j}}^{\dag}\hat{c}_{d_{j+1}}+\hat{c}_{d_{j+1}}^{\dag}\hat{c}_{d_{j}}\right)\nonumber\\
&&+\sum_{i=1}^{N-1}g_{c}\left(\hat{c}_{i}^{\dag}\hat{c}_{i+1}+\hat{c}_{i+1}^{\dag}\hat{c}_{i}\right)+\hat{V}_{1}+\hat{V}_{2},
\end{eqnarray}
with $$\hat{V}_{1}=g_{I}\left(\hat{c}_{l_{n}}^{\dag}\hat{c}_{1}+\hat{c}_{r_{n}}^{\dag}\hat{c}_{N}+\text{H.c.}\right)$$ and $$\hat{V}_{2}=J_{I}\left(|e\rangle\langle g|\hat{c}_{m}+\text{H.c.}\right),$$ where $g_{I}$ represents the register-channel coupling. Hereafter $d$ stands for $\left\{l,r\right\}$. The CCAs are initially prepared in their vacuum states, containing no atom excitation. Then, a single-photon is injected into the left register to have, for example, an arbitrary input state $$|\phi\rangle_{l}=\sum_{j=1}^{n}\alpha_{j}\hat{c}_{l_{j}}^{\dag}|\text{vac}\rangle_{l},$$ where $|\text{vac}\rangle_{l}$ is the vacuum state of the left register. This implies that the dynamics of the system is confined in a single-excitation subspace spanned by the basis vectors
$\left\{|\boldsymbol{d}_{\boldsymbol{j}}\rangle,|\boldsymbol{i}\rangle,|\boldsymbol{e}\rangle\right\}$, where we define
\begin{eqnarray}
|\boldsymbol{d}_{\boldsymbol{j}}\rangle=\hat{c}_{d_{j}}^{\dag}|\text{vac}\rangle|g\rangle,\quad
|\boldsymbol{i}\rangle=\hat{c}_{i}^{\dag}|\text{vac}\rangle|g\rangle,\quad
|\boldsymbol{e}\rangle=|\text{vac}\rangle|e\rangle,\nonumber
\end{eqnarray}
and $|\text{vac}\rangle$ is the vacuum state of the three CCAs. The unitary evolution under $H_{T}$ results in
\begin{equation}
|\varphi\left(t\right)\rangle=\sum_{j=1}^{n}\alpha_{j}\left[f_{d_{j},l_{j}}\left(t\right)|\boldsymbol{d}_{\boldsymbol{j}}\rangle+\sqrt{\epsilon^{d}_{j}}|\boldsymbol{\epsilon}_{\boldsymbol{j}}
^{\boldsymbol{d}}\rangle\right],
\end{equation}
where $$f_{d_{j'},l_{j}}\left(t\right)=\langle \boldsymbol{d}_{\boldsymbol{j'}}|e^{-i\hat{H}_{T}t}|\boldsymbol{l}_{\boldsymbol{j}}\rangle$$ is the transition amplitude of an excitation between the cavities $l_{j}$ and $d_{j'}$, $\epsilon^{d}_{j}=1-|f_{d_{j},l_{j}}\left(t\right)|^{2}$, and $|\boldsymbol{\epsilon}_{\boldsymbol{j}}^{\boldsymbol{d}}\rangle$ is a normalized linear combination of all the basis vectors apart from $|\boldsymbol{d}_{\boldsymbol{j}}\rangle$.

We consider the limit $$\left\{g_{0},g_{I},J_{I}\right\}\ll g_{c},$$ and work perturbatively in $\hat{V}_{1}$ and $\hat{V}_{2}$. Through an orthogonal transformation $\hat{c}_{i}=\sum_{k=1}^{N}\psi_{i,k}\hat{f}_{k}$ with $$\psi_{i,k}=\sqrt{\frac{2}{N+1}}\sin\left(\frac{ik\pi}{N+1}\right),$$ one can find that the bare channel possesses a bosonic spectrum of $\Lambda_{k}=2g_{c}\cos\left[k\pi/\left(N+1\right)\right]$ \cite{nonuniform1,ut}. Consequently, $\hat{V}_{1}$ and $\hat{V}_{2}$ are transformed to $$\hat{V}_{1}=g_{I}\sum_{k=1}^{N}\psi_{1,k}\left[\hat{c}_{l_{n}}^{\dag}\hat{f}_{k}+\left(-1\right)^{k-1}\hat{c}_{r_{n}}^{\dag}\hat{f}_{k}+\text{H.c.}\right]$$ and $$\hat{V}_{2}=J_{I}\sum_{k=1}^{N}\psi_{m,k}\left(|e\rangle\langle g|\hat{f}_{k}+\text{H.c.}\right),$$ respectively. To control the coherent transport of a single-photon, we restrict our attention to odd $N$, which yields the existence of a single zero-energy mode in the bare channel corresponding to $k=z\equiv\left(N+1\right)/2$. Thus the registers and the atom are resonantly coupled to this mode. In the limit of $g_{0}\ll g_{c}$, the width of the energy band of each register, $$\Omega_{0}=|\lambda_{1}-\lambda_{n}|,$$ would be much smaller than the energy gap between the $z$th and ($z\pm1$)th mode of the bare channel, $$\Omega_{1}=|\Lambda_{z\pm1}-\Lambda_{z}|;$$ that is, $\Omega_{0}\ll\Omega_{1}$. In combination with $$\left\{g_{I}\psi_{1,z},J_{I}|\psi_{m,z}|\right\}\ll\Omega_{1},$$ the registers and the atom are significantly detuned from the nonzero-energy modes of the bare channel, thereby neglecting these off-resonant couplings leads to an effective Hamiltonian
\begin{eqnarray}\label{eff_Ham}
\hat{H}_{\text{eff}}&=&\sum_{d=l,r}\sum_{j=1}^{n-1}g_{j}\left(\hat{c}_{d_{j}}^{\dag}\hat{c}_{d_{j+1}}+\hat{c}_{d_{j+1}}^{\dag}\hat{c}_{d_{j}}\right)\nonumber\\
&&+g_{I}\psi_{1,z}\left[\hat{c}_{l_{n}}^{\dag}\hat{f}_{z}+\left(-1\right)^{z-1}\hat{c}_{r_{n}}^{\dag}\hat{f}_{z}+\text{H.c.}\right]\nonumber\\
&&+J_{I}\psi_{m,z}\left(|e\rangle\langle g|\hat{f}_{z}+\text{H.c.}\right)
\end{eqnarray}
as also shown in Fig. \ref{fig1} (a). This dynamics can be used to make a single-photon switch based upon the dressing of the zero-energy mode by the atom.

%%%%%%%%%%%%%%%%%%%%%%%%%%%%%%%%%%%%%%%%%%%%%%%%%%%%%%%%%%%%%%%%%%%%%%%%%%%%%%%%%%%%%%%%%%%
\begin{figure}[!ht]%[tpb]
\begin{center}
\includegraphics[width=7.6cm,angle=0]{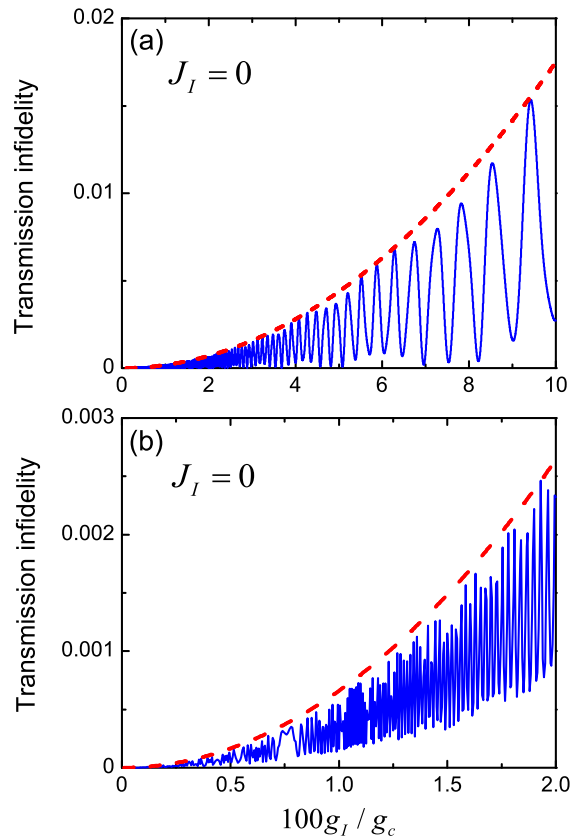}
\caption{(color online) Numerical simulation results of the transmission infidelity $\xi_{r}$ in the uncoupled case ($J_{I}=0$) of either $N=7$, $n=2$ for (a) or $N=101$, $n=10$ for (b). The analytic upper bound is shown by the dashed red curve.}\label{fig2}
\end{center}
\end{figure}
%%%%%%%%%%%%%%%%%%%%%%%%%%%%%%%%%%%%%%%%%%%%%%%%%%%%%%%%%%%%%%%%%%%%%%%%%%%%%%%%%%%%%%%%%%%%%%

If the atom is uncoupled to the cavity ($J_{I}=0$) \cite{atom}, the two spatially-separated registers are coherently coupled by means of the bare channel. It follows, on choosing $g_{I}\psi_{1,z}=g_{n}$ \cite{nonuniform1}, that $$\hat{c}_{l_{j}}^\dag\left(\tau\right)=\left(-1\right)^{n+z-1}\hat{c}_{r_{j}}^\dag$$ for a specific time $\tau=\pi/g_{0}$, which leads to $$f_{r_{j},l_{j}}\left(\tau\right)=\left(-1\right)^{n+z-1}.$$ We therefore have $|\varphi\left(\tau\right)\rangle=\sum_{j=1}^{n}\alpha_{j}|\boldsymbol{r_{j}}\rangle$, implying that the photon is transported from the left register to the right register, and the time evolution is referred to as a swap gate between the two registers [see Fig. \ref{fig1}(b)]. However, the $z$th mode of the bare channel can, in the case when the atom is in the coupled state, be split into a doublet of dressed-states separated by $$\Omega_{2}=2J_{I}|\psi_{m,z}|.$$ Under the assumption that $g_{0}\ll J_{I}$, the two boundary registers are significantly detuned from the two dressed-states if $m$ is odd, and hence, the photon is reflected off the channel, from which the left register is decoupled. In this case, the effective Hamiltonian of Eq. (\ref{eff_Ham}) is reduced to
\begin{eqnarray}
&\hat{H}_{\text{eff}}&=\sum_{j=1}^{n-1}g_{j}\left(\hat{c}_{l_{j}}^{\dag}\hat{c}_{l_{j+1}}+\hat{c}_{l_{j+1}}^{\dag}\hat{c}_{l_{j}}\right)\nonumber\\
&&=\sum_{j,j'=1}^{n}\hat{A}_{j,j'}\hat{c}_{l_{j}}^{\dag}\hat{c}_{l_{j'}},
\end{eqnarray}
where $\hat{A}$ is an $n\times n$ coupling matrix. Furthermore, applying the Heisenberg equations of motion for the operators gives $$\hat{c}_{l_{j}}^{\dag}\left(t\right)=\sum_{j'=1}^{n}\left[\exp{\left(i\hat{A}t\right)}\right]_{j,j'}\hat{c}_{l_{j'}}^{\dag}.$$ Owing to $$\left(\lambda_{q+1}-\lambda_{q}\right)/2g_{0}=1,$$ we find that $$\hat{A}=2g_{0}\hat{P}^{-1}\hat{S}_{x}\hat{P},$$ where $\hat{S}_{x}$ is the $x$ component of a pseudo angular momentum $S=\left(n-1\right)/2$, and $\hat{P}$ is a similarity transformation matrix. In the Schwinger picture \cite{Sch_pic1,Sch_pic2}, $\hat{S}_{x}$ can be expressed in terms of two bosons $\hat{\gamma}_{1}$, $\hat{\gamma}_{2}$, and therefore be thought of as a fictitious Hamiltonian, $$\hat{S}_{x}=\left(\hat{\gamma}_{1}^{\dag}\hat{\gamma}_{2}+\hat{\gamma}_{2}^{\dag}\hat{\gamma}_{1}\right)/2.$$ By calculating the time-evolution operator under this fictitious Hamiltonian, we straightforwardly obtain $$\exp\left(i2g_{0}\hat{S}_{x}\tau\right)\hat{\gamma}_{1}^{\dag}\exp\left(-i2g_{0}\hat{S}_{x}\tau\right)=-\hat{\gamma}_{1}^{\dag},$$ and $$\exp\left(i2g_{0}\hat{S}_{x}\tau\right)\hat{\gamma}_{2}^{\dag}\exp\left(-i2g_{0}\hat{S}_{x}\tau\right)=-\hat{\gamma}_{2}^{\dag},$$ such that $$\exp{\left(i\hat{A}\tau\right)}=(-1)^{n-1}\hat{I},$$ yielding $$\hat{c}_{l_{j}}^{\dag}\left(\tau\right)=\left(-1\right)^{n-1}\hat{c}_{l_{j}}^{\dag}.$$ The resulting transition amplitude is $$f_{l_{j},l_{j}}\left(\tau\right)=(-1)^{n-1}.$$ The final state of the system then becomes $|\varphi\left(\tau\right)\rangle=|\varphi\left(0\right)\rangle$, and thus the quantum state of the input photon remains unchanged after time evolution of functioning as an identity operation [see Fig. \ref{fig1}(c)].

%%%%%%%%%%%%%%%%%%%%%%%%%%%%%%%%%%%%%%%%%%%%%%%%%%%%%%%%%%%%%%%%%%%%%%%%%%%%%%%%%%%%%%%%%%%%%%%%%%%%%%%%%%%%%%%%%%%%%%%%%%%%%%%%%%%%%%%%%%%%%%%%%%%%%%%%%%%%%%%%%%%%%%%%%%%%%%%%%%%%%%%%%%%%%%%
\section{Leakage of quantum information}
%%%%%%%%%%%%%%%%%%%%%%%%%%%%%%%%%%%%%%%%%%%%%%%%%%%%%%%%%%%%%%%%%%%%%%%%%%%%%%%%%%%%%%%%%%%
\begin{figure}[!ht]%[tpb]
\begin{center}
\includegraphics[width=7.6cm,angle=0]{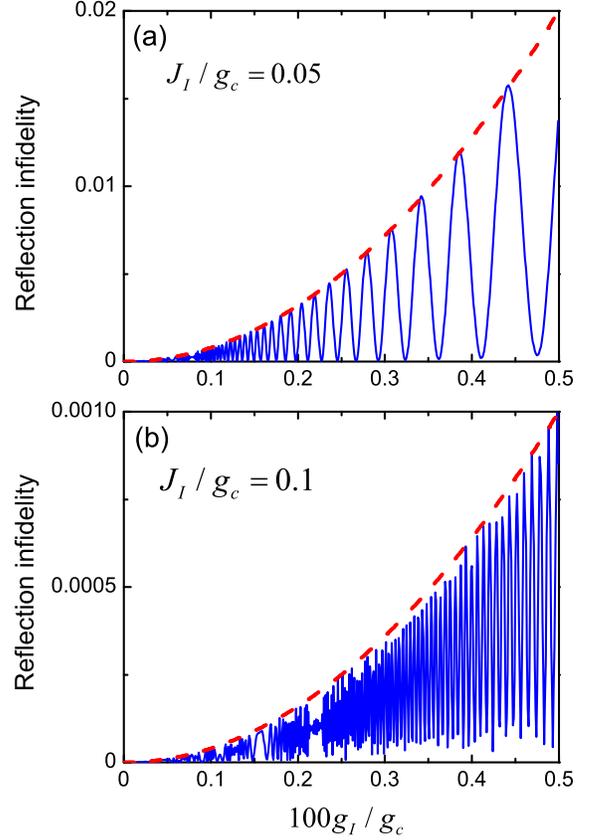}
\caption{(color online) Numerical simulation results of the reflection infidelity $\xi_{l}$ in the coupled case of either $N=7$, $n=2$, $J_{I}/g_{c}=0.05$ for (a) or $N=101$, $n=10$, $J_{I}/g_{c}=0.1$ for (b). The analytic upper bound is shown by the dashed red curve. Here, we choose $m=3$.}\label{fig3}
\end{center}
\end{figure}
%%%%%%%%%%%%%%%%%%%%%%%%%%%%%%%%%%%%%%%%%%%%%%%%%%%%%%%%%%%%%%%%%%%%%%%%%%%%%%%%%%%%%%%%%%%%%%

Having explicitly demonstrated eigenmode-mediated single-photon transport, we now calculate its leakage of quantum information by making use of perturbation theory. Such a leakage arises only from the off-resonant couplings between the registers and the channel. Upon performing a first-order perturbative treatment, we find that the leakage of quantum information results from the two cavities coupled directly to the intermediate channel. Specifically, in the uncoupled case, the full dynamics can be mapped onto an effective photon transport channel being perturbatively coupled to a finite bosonic environment, whose Hamiltonian is $$\hat{H}_{z}=\sum_{k\neq z}\Lambda_{k}\hat{f}_{k}^{\dag}\hat{f}_{k}.$$ The interaction between them is $$\hat{V}_{z}=g_{I}\sum_{k\neq z}\psi_{1,k}\left[\hat{c}_{l_{n}}^{\dag}\hat{f}_{k}+\left(-1\right)^{k-1}\hat{c}_{r_{n}}^{\dag}\hat{f}_{k}+\text{H.c.}\right].$$ Up to second order, $\epsilon_{j}^{r}$ is modified as $$\epsilon_{j}^{r}\simeq4\Delta_{r}\delta_{n,j},$$ where
\begin{equation}\label{eq:deltar}
\Delta_{r}=\sum_{k<z}\Delta_{k}^{r}\left[1-\left(-1\right)^{n+k+z}\cos\left(\Lambda_{k}\tau\right)\right]
\end{equation}
and $$\Delta^{r}_{k}=\left(g_{I}\psi_{1,k}/\Lambda_{k}\right)^{2}.$$ In the coupled case, the zero-energy mode of the bare channel is the only state that is dressed by the atom owing to $J_{I}\ll g_{c}$.  The boundary registers are thus coupled to the two dressed-states in addition to the bosonic environment; however, the coupling to this environment can be neglected so long as $g_{0}\ll J_{I}$. With a similar perturbative treatment as before, $\epsilon_{j}^{l}$ is given by $$\epsilon_{j}^{l}\simeq4\Delta_{l}\delta_{n,j},$$ where
\begin{equation}\label{eq:deltal}
\Delta_{l}=\Delta_{z}^{l}\left[1-\left(-1\right)^{n-1}\cos\left(J_{I}\psi_{m,z}\tau\right)\right]
\end{equation}
and $$2\Delta_{z}^{l}=\left(g_{I}\psi_{1,z}/J_{I}\psi_{m,z}\right)^{2}.$$ Observing these exhibits that encoding quantum information into the cavities between $d_{1}$ and $d_{n-1}$ could be more efficient.

In order to quantify quantum information leaking into the off-resonant modes of the intermediate channel, we need to employ two average fidelities, the reflection fidelity $$F_{l}=\int\!\! d\phi\;\langle\phi|\hat{\rho}_{l}\left(\tau\right)|\phi\rangle$$ and the transmission fidelity $$F_{r}=\int\!\! d\phi\;\langle\phi|\hat{\rho}_{r}\left(\tau\right)|\phi\rangle.$$ Here, $\hat{\rho}_{l/r}\left(\tau\right)$ is the output reduced density matrix of the left or right register, the integration is over all input pure states and $\int\!\! d\phi$ is normalized to unity. The fidelity $F_{d}$ (in combination with $F_{l}$ and $F_{r}$) can, after a straightforward calculation, be expressed in terms of the transition amplitudes,
\begin{equation}\label{eq:avera_fidel}
F_{d}=\frac{1}{n\left(n+1\right)}\sum_{j,j'=1}^{n}\left[|f_{d_{j'},l_{j}}\left(\tau\right)|^{2}+f_{d_{j},l_{j}}\left(\tau\right)f_{d_{j'},l_{j'}}^{*}\left(\tau\right)\right].
\end{equation}
To demonstrate our theoretical results, we numerically simulate the transmission infidelity $$\xi_{r}=1-F_{r}$$ (see Fig. \ref{fig2}) and the reflection infidelity $$\xi_{l}=1-F_{l}$$ (see Fig. \ref{fig3}) for the $N=7$, $n=2$ and $N=101$, $n=10$ cases, as two examples. Specifically, in finite channels of fixed length, the infidelity $\xi_{d}$ is plotted as a function of $g_{I}/g_{c}$ along with an analytic upper bound. Working within the weak-coupling limit, $\xi_{d}$ can be analytically expressed as $\xi_{d}\simeq2\Delta_{d}$, and has the upper bound
\begin{equation}\label{eq:upper_bounds}
\xi_{d}\leq\frac{8}{n}\left(\Delta_{z}^{l}\delta_{l,d}+\sum_{k<z}\Delta_{k}^{r}\delta_{r,d}\right).
\end{equation}
This upper bound is in excellent agreement with the numerical results, shown in Figs. \ref{fig2} and \ref{fig3} \cite{fig23}. In addition, we find that decreasing $g_{I}/g_{c}$ can suppress the leakage of quantum information, so $\xi_{d}$ can in principle be made arbitrarily small.

\begin{figure}[htbp]%[tpb]
\begin{center}
\includegraphics[width=7.3cm,angle=0]{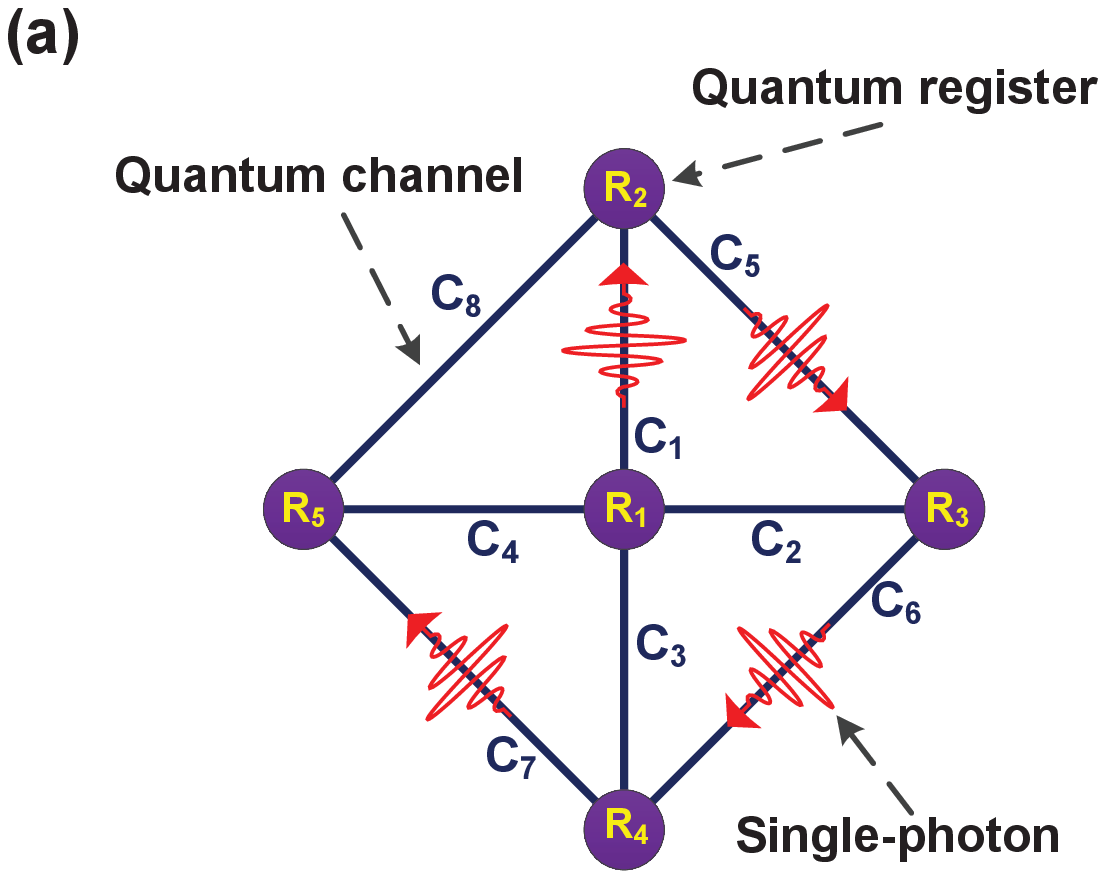}
\includegraphics[width=8.5cm,angle=0]{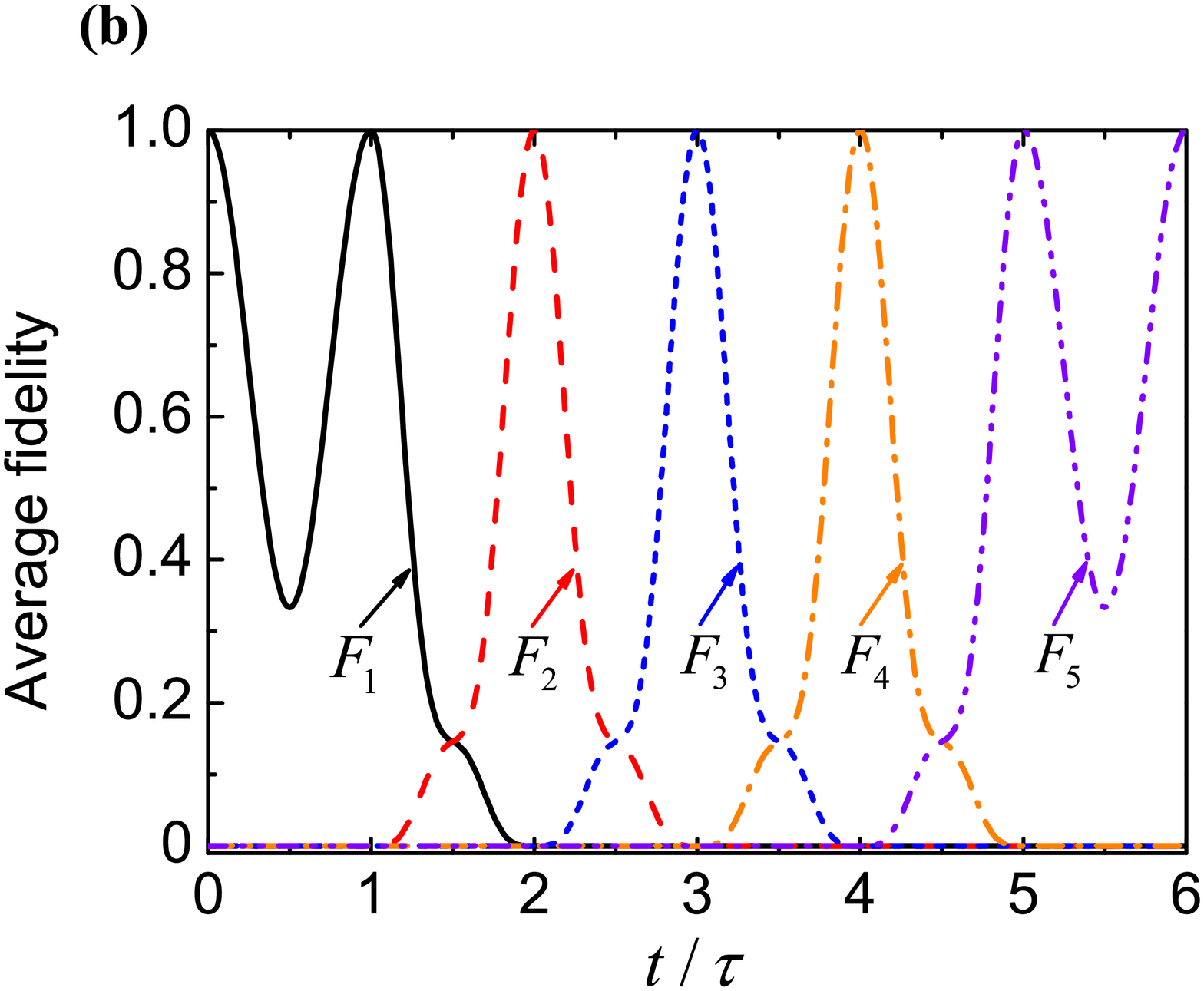}
\caption{(color online) (a) Schematic illustration of a network which is made up of five registers and eight channels. Each register is coupled to at least three channels. Depending upon the atomic state, a single-photon can be stored in one of the registers or transported between them as required. (b) The average fidelities plotted as functions of the evolution time for a single-photon travelling along the network in (a). We choose that all the atoms are in the coupled states during the time intervals $[0,\tau]$ and $(5\tau,6\tau]$, with $J_{I}/g_{c}=0.05$; while the atoms in the channels $C_{1}$, $C_{5}$, $C_{6}$ and $C_{7}$ are uncoupled during the time intervals $(\tau,2\tau]$, $(2\tau,3\tau]$, $(3\tau,4\tau]$ and $(4\tau,5\tau]$, respectively. The solid black curve in (b) corresponds to $F_{1}$, the dashed red curve to $F_{2}$, the short dashed blue curve to $F_{3}$, the dashed-dotted orange curve to $F_{4}$, and the dashed-double dotted violet curve to $F_{5}$. Here, $g_{I}/g_{c}=0.0001$, $N=7$, $n=2$ and $m=3$.}\label{fig4}
\end{center}
\end{figure}

\section{Extensions}
In direct analogy to a classical computer, the potential power of a quantum computer exponentially increases with the number of qubits, but increasing arbitrarily the number of qubits is not easy to achieve. One approach to addressing this challenge is to envision a quantum computer containing a number of quantum registers \cite{network}, so the study of the multi-register setups is important for making a powerful future quantum computer. While we have focused on the two-register case, the extension to the multi-register networks is directly analogous. In such networks, the registers and the channels are the same as mentioned in the description above, except that a single register needs to be coupled to multiple channels. The couplings of the channels to the registers and to the atoms are also chosen as before. The state of all the atoms in the coupled state decouples all the registers from the channels, thus, quantum information will be stored in the independent registers. However, in the situation where one of the atoms is uncoupled,  the corresponding bare channel coherently couples two distant registers which are still decoupled from other channels, and therefore information transport will be reliably achieved between them. Together with individually addressable atoms, quantum information can be redirected from one register to another, in direct analogy to a quantum routing function. For simplicity, let us consider a specific network of five registers $R_{1},\cdots,R_{5}$ and eight channels $C_{1},\cdots,C_{8}$, and demonstrate a single-photon travelling along the path  $R_{1}\rightarrow R_{2}\rightarrow R_{3}\rightarrow R_{4}\rightarrow R_{5}$, shown in Fig. \ref{fig4}(a). Suppose now that a single-photon is initially prepared in the register $R_{1}$ with an arbitrary input state. To confirm this travel, we numerically simulate the average fidelity $F_{\theta}$ ($\theta=1,\cdots,5$) between the input state of the register $R_{1}$ and the output state of the register $R_{\theta}$ [see Fig. \ref{fig4}(b)]. These numerical results show that the controllable single-photon transport in a network can be achieved with very high fidelity. Despite the fact that we elucidate only one of the paths in a simple network, in principle, our method can enable any arbitrary path and more complex networks.

\section{Conclusions}
We have proposed and analyzed single-photon controllable transport using a 1D CCA to coherently couple two identical spatially-separated 1D CCAs, and a two-level atom, to control the transport of single-photons. We study the pure Hamiltonian evolution in this hybrid system. In the case when the atom is absent, a single-photon with an arbitrary unknown quantum state (for example, initially in the left CCA) will be transported to the right CCA, with a transmission fidelity arbitrarily close to unity. On the contrary, as a result of the coupling of the atom to the intermediate CCA, this single-photon will be reflected back into the left CCA and leave its quantum state unchanged, with a reflection fidelity also arbitrarily close to unity. The approach can also be directly generalized to multi-register quantum networks, and thus due to its scalability, applied to realize quantum information processing devices. It should be noted that, in the no-atom case, this method allows for arbitrary multi-photon state coherent transport through the intermediate CCA, even in a thermal equilibrium state.  The proposed setup can be examined in the context of circuit quantum electrodynamics with superconducting circuits. For example, superconducting qubits act as two-level atoms and transmission line resonators bahave as cavities. In this situation, two nearest-neighbor resonators can be straightforwardly connected via a capacitor. The coupling strength could experimentally reach $2\pi\times 31$ MHz \cite{inter_resonator_coupling} and in fact this reachable strength can be markedly larger by increasing the capacitance. Moreover, the coupling between single superconducting qubits and transmission line resonators has also been implemented in the strong-coupling regime and even in the ultrastrong-coupling regime, with strengths of up to $2\pi\times 107$ MHz \cite{strong} and $12\%$ of the cavity frequency \cite{ustrong}, respectively. Hence, our theoretical model seems to be experimentally accessible using current technologies. While we have chosen to focus on the special case of a CCA system, this framework can be employed to achieve controllable quantum state transfer in a wide range of systems, including, for example, coupled quantum spin chains.

%%%%%%%%%%%%%%%%%%%%%%%%%%%%%%%%%%%%%%%%%%%%%%%%%%%%%%%%%%%%%%%%%%%%%%%%%%%%%%%%%%%%%%%%%%%%%%%%%%%%%%%%%%%%%%%%%%%%%%%%%%%%%%%%%%%%%%%%%%%%%%%%%%%%%%%%%%%%%%%%%%%%%%%%%%%%%%%%%
\section{acknowledgments}
WQ is supported by the Basic Research Fund of the Beijing Institute of Technology under Grant No. 20141842005. FN is supported by the RIKEN iTHES Project, MURI Center for Dynamic Magneto-Optics via the AFOSR Award No. FA9550-14-1-0040, the Impact Program of JST, a Grant-in-Aid for Scientific Research (A).

%%%%%%%%%%%%%%%%%%%%%%%%%%%%%%%%%%%%%%%%%%%%%%%%%%%%%%%%%%%%%%%%%%%%%%%%%%%%%%%%%%%%%%%%%%%%%%%%%%%%%%%%%%%%%%%%%%%%%%%%%%%%%%%%%%%%%%%%%%%%%%%%%%%%%%%%%%

%%%%%%%%%%%%%%%%%%%%%%%%%%%%%%%%%%%%%%%%%%%%%%%%%%%%%%%%%%%%%%%%%%%%%%%%%%%%%%%%%%%%%%%%%%%%%%%%%%%%%%%%%%%%%%%%%%%%%%%%%%%%%%%%%%%%%%%%%%%%%%%%%%%%%%%%%%%%%%%%%%%%%%%%%

\begin{thebibliography}{99}
\bibitem{atom} In fact, we can introduce an auxiliary atomic state $|\tilde{g}\rangle$ in addition to the states $|g\rangle$ and $|e\rangle$. The states $|g\rangle$ and $|\tilde{g}\rangle$ belong to the ground state manifold. We assume that $|\tilde{g}\rangle$ is completely uncoupled to the cavity mode, so that the interaction $V_{2}$ does not play a role in Eq. (\ref{T_Hami}) if the atom is in $|\tilde{g}\rangle$, thus equivalent to having $J_{I}=0$.
\bibitem{fig23} In Figs. \ref{fig2} and \ref{fig3}, the infidelities and the upper bounds are plotted according to Eqs. (\ref{eq:avera_fidel}) and (\ref{eq:upper_bounds}), respectively. The fidelity in Eq. (\ref{eq:avera_fidel}) has been averaged over all possible input pure states of the left register, so the plotted infidelities and upper bounds are the input-state-averaged quantities. Furthermore, we plot these quantities at a specific evolution time such that the quantum state of either the transmitted or reflected single-photon remains unchanged.
\end{thebibliography}
\end{document}